# Giant resonant radiative heat transfer between nanoparticles

Yong Zhang[1,2], Hong-Liang Yi[1,2,*], He-Ping Tan[1,2], and Mauro Antezza[3,4,†]

[1]*School of Energy Science and Engineering, Harbin Institute of Technology, Harbin 150001, P. R. China*
[2]*Key Laboratory of Aerospace Thermophysics, Ministry of Industry and Information Technology, Harbin 150001, P. R. China*
[3]*Laboratoire Charles Coulomb (L2C), UMR 5221 CNRS-Université de Montpellier, F- 34095 Montpellier, France*
[4]*Institut Universitaire de France, 1 rue Descartes, F-75231 Paris, France*

We show that periodic multilayered structures allow to drastically enhance near-field radiative heat transfer between nanoparticles. In particular, when the two nanoparticles are placed on each side of the multilayered structure, at the same interparticle distance the resulting heat transfer is more than five orders of magnitude higher than that in the absence of the multilayered structure. This enhancement takes place in a broad range of distances and is due to the fact that the intermediate multilayered structure supports hyperbolic phonon polaritons with the key feature that the edge frequencies of the Type I and Type II Reststrahlen bands coincide with each other at a value extremely close to the particle resonance. This allow a very high-$k$ evanescent modes resonating with the nanoparticles. Our predictions can be relevant for effective managing of energy at the nano-scale.

Since the pioneering work of Polder and van Hove [1] it is well known that when two objects are brought in proximity to each other (i.e., in the near-field regime), the radiative heat transfer (RHT) between them may be significantly enhanced [2]. This is caused by the tunneling effect of evanescent modes, as surface plasmon polaritons (SPPs) or surface phonon polaritons (SPhPs) [3–9]. This effect has several implications in various technologies for near-field energy conversion [10,11] and data storage [12] as well as active thermal management [13] at nanoscale with, transistors [14,15], thermal rectifiers [16–19], memories [20,21].

For these reasons, huge efforts have been done to find new configurations where this effect can be further enhanced or modulated. Recently, it has been shown that, due to the presence of many-body interactions, strong exaltation effects of heat flux are possible [22–26]. Focus has been done on the RHT between two particles in the presence of one or two plates [27–31], where the two particles lie on the same sides of the plate, and a significant amplification of RHT is shown at a long-range distance. Note that in the context of Forster/Resonant energy transfer [32], the transmission configurations where the two dipoles are placed on each side of a slab with finite thickness have also been adopted theoretically and experimentally in [33,34]. Among the possible strategies, multilayers have been extensively studied because mutual interactions of surface polaritons at multiple interfaces inside multilayers [35–38] provide exotic features including tuning of near-field thermal radiation.

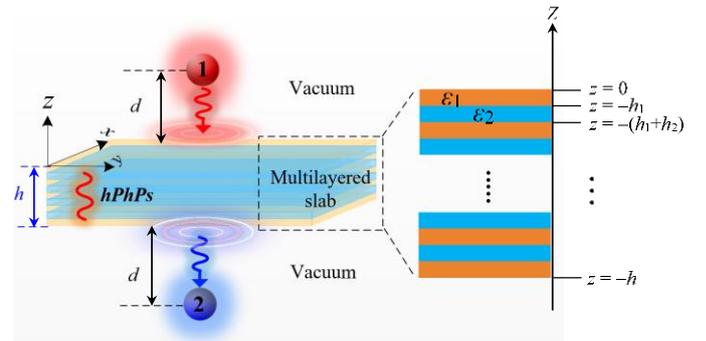

FIG. 1 Scheme of the RHT between two nanoparticles in the presence of a multilayered slab. The nanoparticles lie on the opposite sides of the intermediate structure with a total thickness of $h$. The top surface of the slab coincides with the coordinate origin. The periodic multilayered structure consists of multiple layers of alternating materials.

In this Communication, we study the RHT between nanoparticles which are separated by a multilayered structure, as shown in Fig. 1. Remarkably, we show that the particle to particle heat transfer can be up to more than five orders of magnitude higher than that in the absence of the intermediate structure. We explain this effect in terms of a new hyperbolic channel having peculiar features.

We consider two particles which are isotropic spherical with radius $R = 5$nm and modeled as simple radiating dipoles, which is valid when the center-to-center distance $L > 3R$ [27–30]. We choose the nanoparticles made of silicon carbide (SiC), with its dielectric function described by the Drude-Lorentz model [39]: $\varepsilon(\omega) = \varepsilon_\infty \left(\omega_L^2 - \omega^2 - i\Gamma\omega\right)/\left(\omega_T^2 - \omega^2 - i\Gamma\omega\right)$ with high-frequency dielectric constant $\varepsilon_\infty = 6.7$, longitudinal optical frequency $\omega_L = 1.83 \times 10^{14}$ rad/s, transverse optical frequency $\omega_T = 1.49 \times 10^{14}$ rad/s, and damping $\Gamma = 8.97 \times 10^{11}$ rad/s. The periodic multilayered structure consists of multiple layers of alternating materials, namely film 1 and film 2. Each material has an individual thickness $h_i$ ($i = 1, 2$) and if we have $N$ layers the total thickness of the complete multilayered structure is $h = [(N+1)h_1 + (N-1)h_2]/2$. We choose film 1 made by SiC and film 2 made by a polar dielectric NaBr, with its dielectric function given as [40]: $\varepsilon_2(\omega) = \varepsilon_\infty \left(\omega_L^2 - \omega_T^2\right)/\omega_T^2 - \omega^2 + i\Gamma\omega$, where $\varepsilon_\infty = 2.6$, $\omega_L = 0.39 \times 10^{14}$ rad/s, $\omega_T = 0.25 \times 10^{14}$ rad/s and $\Gamma = 2.6 \times 10^{11}$ rad/s. The coordinate system is defined such that the $x$–$y$-plane coincides with the surface of the topmost layer and the $z$-direction is orthogonal to the surface. Both the two nanoparticles are put in proximity to the surface at a particle-surface distance of $d$. The vertical positions of the two particles are thus given by $z_1 = d$ and $z_2 = -h-d$, respectively.

We assume that the whole system is initially thermalized at a given temperature $T$, and we then give the top particle a tiny temperature increment $\Delta T$. The only net heat exchanged by bottom particle is thus with the top one. When $\Delta T \to 0$, we get the RHT conductance $\Phi$ between these two particles which is the quantity we are going to discuss in this Communication. Note that the heat exchange between the top article and the intermediate structure will not affect the energy transferred between the two particles.

According to the framework of fluctuational electrodynamics (FE), the conductance $\Phi$ between two identical nanoparticles at a temperature of $T$ can be conveniently expressed in terms of the Green's function (GF) describing the system as [28],

$$\Phi = 4\int_0^{+\infty} \frac{d\omega}{2\pi} \hbar\omega n'(\omega,T) k_0^4 \left[\text{Im}(\alpha)\right]^2 \text{Tr}\left[\mathbb{G}\mathbb{G}^*\right], \quad (1)$$

where $k_0 = \omega/c$ and $\alpha$ denoting the free-space wave-vector and the particle's electric frequency-dependent polarizability, respectively. In the limit $R \ll \delta$ (with $\delta$ being the skin depth of the given material), $\alpha$ is written in the well-known Clausius-Mossoti form $\alpha(\omega) = 4\pi R^3 \left[(\varepsilon(\omega)-1)/(\varepsilon(\omega)+2)\right]$. Notice that the expression of $\alpha$ predicts a nanoparticle resonance frequency $\omega_{np}$ corresponding asymptotically to the condition $\varepsilon(\omega) + 2 = 0$, which for SiC gives $\omega_{np} = 1.756 \times 10^{14}$ rad/s. $n'(\omega,T)$ denotes the derivative with respect to $T$ of the Bose-Einstein distribution $n(\omega,T) = \left(\exp(\hbar\omega/k_B T) - 1\right)^{-1}$. $\mathbb{G}$ denotes the GF. For the two isolates particles, the well-known free-space GF reads,

$$\mathbb{G} = \mathbf{G}_0 = \frac{e^{ik_0 L}}{4\pi k_0^2 L^3}\begin{pmatrix} a & 0 & 0 \\ 0 & b & 0 \\ 0 & 0 & b \end{pmatrix}, \quad (2)$$

where $a = 2 - 2ik_0 L$ and $b = k_0^2 L^2 + ik_0 L - 1$. In the presence of an intermediate multilayered slab, Eq. (2) is replaced by the transmitted GF [33,41],

$$\mathbb{G} = \mathbf{G}_{tr} = \frac{i}{4\pi}\int_0^{+\infty} \left[t_s \mathbf{M}_{tr}^s + t_p \mathbf{M}_{tr}^p\right] e^{i[-k_{z0}(z_2+h)+k_{z0}z_1]} k_\rho dk_\rho, \quad (3)$$

where $k_\rho$ and $k_{z0} = \sqrt{k_0^2 - k_\rho^2}$ are the lateral wave-vector to the surface and the $z$ component of the wave-vector in vacuum, respectively. $t_s$ and $t_p$ are the Fresnel transmission coefficients of the total structure associated with the two polarizations. For the exact solution, they take the form of $t = \left(t_{01}t_{12}e^{ik_{z,1}h_1}\right)/\left(1 - r_{10}r_{12}e^{i2k_{z,1}h_1}\right)$, where we denote 0, 1, and 2 the vacuum region above the film 1, the film 1 and the region below the film 1, respectively. One can refer to Refs. [35,42] for the detailed calculation method of $t$. The transmission matrices $\mathbf{M}_{\mathrm{tr}}^{s}$ and $\mathbf{M}_{\mathrm{tr}}^{p}$ are defined as,

$$\mathbf{M}_{\mathrm{tr}}^{s} = \frac{1}{k_{z1}} \begin{bmatrix} \frac{1}{2}A_1 & 0 & 0 \\ 0 & \frac{1}{2}A_2 & 0 \\ 0 & 0 & 0 \end{bmatrix} \quad (4a)$$

$$\mathbf{M}_{\mathrm{tr}}^{p} = \frac{1}{k_0^2} \begin{bmatrix} \frac{k_z}{2}A_2 & 0 & ik_\rho J_1(k_\rho d_x) \\ 0 & \frac{k_z}{2}A_1 & 0 \\ ik_\rho J_1(k_\rho d_x) & 0 & J_0(k_\rho d_x)k_\rho^2/k_z \end{bmatrix} \quad (4b)$$

where $J_n$ is the cylindrical Bessel function of order $n$. $A_1 = J_0(k_\rho d_x) + J_2(k_\rho d_x)$, $A_2 = J_0(k_\rho d_x) - J_2(k_\rho d_x)$ and $d_x$ is the lateral center-to-center distance between the two nanoparticles along the $x$-axis. Note that by setting the transmission coefficients to one, viz., $t_s = t_p = 1$, the transmitted GF in Eq. (3) degrades to the free-space GF, viz. $\mathbf{G}_0$. When both the two particles locate in the upper half-space, the total GF is the sum of the free space GF and the reflected GF $\mathbf{G}_{\mathrm{ref}}$, which has already been discussed in Ref. [27,28].

Let us start the discussion of the results by illustrating the main finding of our work. The two particles are put in proximity to the intermediate multilayered slab at a distance of $d = 50$nm, while the total thickness of the multilayered varies from 100nm to 100μm. In Fig. 2(a), we plot the conductance $\Phi$ at 300K between the two SiC nanoparticles for a layer number of $N = 3$, 11 and 21. The ratio $R$ between $\Phi$ and the conductance in the absence of the structure $\Phi_0$ is shown in Fig. 2(b). Meanwhile, for comparison we also present the results for the single SiC slab and for the reflection model from Ref. [28].

For a single SiC slab ($N = 1$), due to the coupling of the evanescent waves on the up and down interfaces we can also observe an enhancement of RHT. However, with an increase in the thickness the coupling effect of surface waves fade gradually, and eventually vanishes when $h = 4.0$μm. As we adopt the multilayered structures ($N > 1$), a large enhancement RHT is achieved. More specifically,

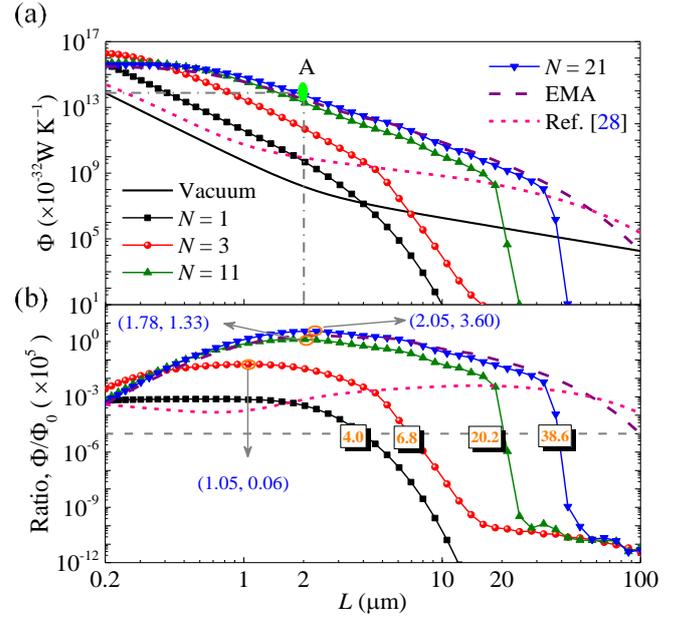

FIG. 2 (a) RHT conductance $\Phi$ between the two particles as a function of the interparticle distance $L$ ($2d + h$). (b) The ratio $R$ between $\Phi$ and the conductance in the absence of the structure $\Phi_0$. The temperature is $T = 300$ K. The particles are placed in proximity to the surface at a distance of $d = 50$ nm. $h$ varies from 100nm to 100μm. Point A in panel (a) denotes the conductance at 2 μm which equals to that at 0.2μm in the absence of the intermediate structure. All the slabs are of equal thickness.

as shown in Fig. 2(b) we observe maximum amplification with values of 6000, 133000, 361000 at thicknesses of 0.95, 1.68 and 1.95μm for $N = 3$, 11 and 21, respectively. That is to say, at the same interparticle distance the resulting heat transfer in the presence of the multilayered structure with $N = 21$ is more than five orders of magnitude higher than that in the absence of the multilayered structure. Furthermore, the amplification persists to large distances of 6.8, 20.2 and 38.6μm, indicating a great enhancement RHT from near-field to far-field. In another view of point, in Fig. 2(a) we denote point A the conductance at 2 μm which equals to that at 0.2μm in the absence of the intermediate structure, which indicates that the same amount of heat flux can be achieved at a long distance as in the near-field regime. This means that the intermediate structure could act as a passive amplifier for the heat signal detection.

Since for a large thickness evanescent waves between the adjacent interfaces could not couple with each other anymore, the RHT from the top particle is invisible to the other side of the slab, hence a suppression of RHT as depicted in Fig. 2. When the two particles locate above a SiC bulk, assisted by the propagation of the surface waves an amplification of RHT is observed [28]. Although both the direct and reflected RHT between the two particles are included, the maximum amplification is 400 at 20μm which is three orders of magnitude less than those with our multilayered structure. This confirms the superiority of our transmission configuration with an intermediate multilayered structure. Note that the above results are obtained exactly based on the multilayered transmission coefficients. In the periodic multilayered media as considered here, the periodicity is far smaller as compared to the relevant thermal wavelength at the chosen temperature. The effective medium approach (EMA) can be used to describe it as a homogenous but uniaxial material. The in-plane and our-of-plane permittivities are thus approximately given by $\varepsilon_\parallel = f\varepsilon_1 + (1-f)\varepsilon_2$ and $\varepsilon_\perp = 1/(f/\varepsilon_1 + (1-f)/\varepsilon_2)$, where $f$ is the filling ratio of the SiC defined as $h_1/(h_1+h_2)$. The EMA results for $f = 0.5$ are presented in Fig. 2 for $N \rightarrow \infty$, which agree well with those of $N = 21$ as $L < 40$μm, and predict an enhancement of RHT even when $L$ is up to 100μm.

The origin of this striking enhancement in RHT conductance can be understood with an analysis of optical property of the multilayered structure by EMA (for more details, see Ref. [43]) and the frequency-wave-vector dependence of the transmission coefficient $t_p$. In Fig. 3(a) we plot the dielectric function for $f = 0.5$ by EMA. We observe two Reststrahlen bands, viz., Type I band at $[1.7424–1.826] \times 10^{14}$ rad/s, and Type II band at $[1.495–1.742] \times 10^{14}$ rad/s, indicating that the multilayered structure supports the hyperbolic phonon polaritons (hPhPs). Interestingly, the edge frequencies of these two bands connect with each other at $\omega_h = 1.742 \times 10^{14}$ rad/s. With $h = 1$μm and $N = 101$, Im($t_p$) distributions by the exact model are plotted in Fig. 3(b). Remarkably, we notice that both the slopes of the bright bands in these two zones approach to zero around the $1.742 \times 10^{14}$ rad/s, indicating very-large-momentum extraordinary rays, viz., high-$k$ evanescent modes, at a frequency extremely close to the SiC nanoparticle resonance at $\omega_{np}$. We stress that the frequency-matched high-$k$ evanescent modes are indeed responsible for the giant enhancement of RHT between nanoparticles. To identify the hPhPs more clearly, we plot the dispersion lines obtained from
$$k_\rho = (-\delta/h)\left[n\pi + \arctan\left(\omega/(\delta\omega\varepsilon_\perp)\right) + \arctan\left(1/(\delta\varepsilon_\perp)\right)\right]$$
[35] in which $\delta = -i\sqrt{\varepsilon_\parallel/\varepsilon_\perp}$. Series of bright bands with strong reflection appear in the two spectral ranges and are in good agreement with the bright bands in Im($t_p$). This is also the evidence of the validity of our exact approach.

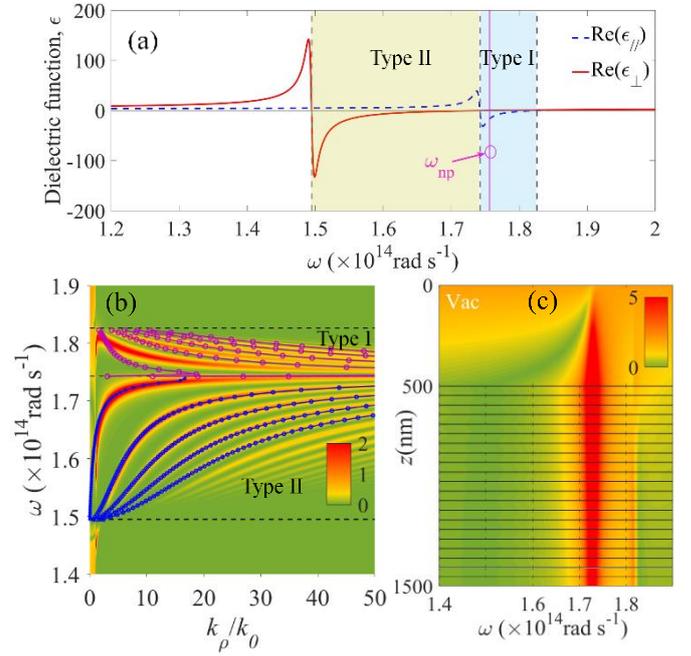

FIG. 3 (a) Dielectric function of the multilayered structure by EMA. Types I and II Reststrahlen bands are shaded in yellow and blue, respectively. (b) Distributions of Im($t_p$) obtained by the exact model for $N = 21$ and $f = 0.5$. The purple and blue dotted lines mark the dispersion relations for Type I and Type II hPhPs, respectively. (c) The absolute $\vec{E}_x$-field amplitudes as a function of $\omega$ and $z$-position for the exact model. $h = 1$μm.

In addition to the transmission coefficient, we adopt the generalized 4 × 4 matrix formalism [44,45] to calculate the electric field distributions. SPhPs are excited in the Otto geometry (for more details, see Ref. [43]). The absolute $\vec{E}_x$-field amplitudes are shown as a function of frequency and $z$-position for the exact model in Fig. 3(b). Remarkably, we see that this field enhancement peaks at $1.742 \times 10^{14}$ rad/s penetrates throughout the multilayered structure, indicating a fluent transmission channel for the evanescent waves around the particle resonance at $\omega_{np}$.

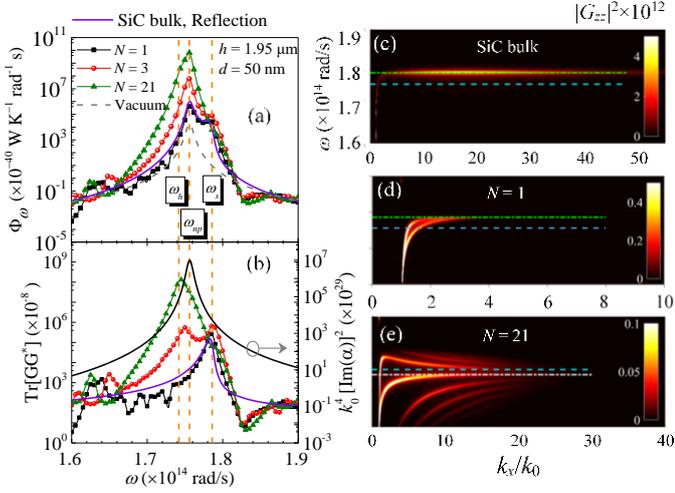

FIG. 4 (a) Spectral conductance $\Phi_\omega$. (b) The trace of the GF and the polarizability of the SiC particles. The absolute square of the $zz$ element $|G_{zz}|^2$ of (c) the reflection GF for the two positions separated at $L = 2.05$μm above the SiC bulk, and the transmission GF for the two positions separate by a multilayer structure with (d) $N = 1$, and (e) $N = 21$. $h$ is 1.95μm for the transmission model and $d = 50$nm. The three vertical yellow dotted lines in panels (a) and (b) mark the three resonances at $\omega_h$, $\omega_{np}$, and $\omega_s$, respectively. And also we denote them as horizontal white, blue and pink dotted lines in panels (c)-(e).

To further interpret the underlying physics of the above results and confirm that the frequency-matched high-$k$ modes are indeed responsible for the giant enhancement of RHT, we show the spectral conductance $\Phi_\omega$ in Fig. 4(a). For the transmission configuration, the total thickness $h$ is chosen as 1.95μm. We note that the SPhPs of the single SiC slab or the SiC bulk exhibit resonance at $\omega_s = 1.786\times 10^{14}$ rad/s corresponding asymptotically to the condition $\varepsilon(\omega) + 1 = 0$. Now we have three characteristic frequencies in this system, namely, $\omega_h$, $\omega_{np}$, and $\omega_s$. Notice that the higher of the $\Phi_\omega$ at $\omega_{np}$ is, the larger total conductance is obtained as shown in Fig. 2(a). While for the single slab or the bulk, a second peak with small value emerges at $\omega_s$. However, the contribution from this peak to the RHT between nanoparticles is limited. Physically, one can expect that the heat flux results from the interaction between the waves emitted from the particles and those from the slab or surface. Mathematically, the formulation for the conductance expressed in Eq. (1) explicitly includes these two waves' characteristics in the polarizability $\alpha$ and the GF, respectively. More specifically, we separate the conductance into two quantities, viz., $k_0^4 \left[ \text{Im}(\alpha) \right]^2$ and $\text{Tr}\left[ \mathcal{G}\mathcal{G}^* \right]$ as shown in Fig. 4(b). For the bulk or single film, the trace of the GF only peaks at $\omega_s$. However the value of the polarizability at $\omega_s$ is far lower than that at $\omega_{np}$. Hence, as we multiply these two quantities, the amplification contributed from the peak on the GF is weakened. Meanwhile, due to the very small value of $\text{Tr}\left[ \mathcal{G}\mathcal{G}^* \right]$ at $\omega_{np}$, the particle resonance's contribution to the RHT is also limited. As a result, no matter the transmission or the reflection configuration is used, the amplification for the single SiC is very limited. As the multilayered structure with $N = 3$ is used, we see in Fig. 4(a) that an additional peak emerges at a frequency very close to $\omega_{np}$ on the $\text{Tr}\left[ \mathcal{G}\mathcal{G}^* \right]$, and the peak at $\omega_s$ increases to a bigger value than the case with the bulk or single film. By further increasing $N$ to 21, due to the increasingly prominent effect of hPhPs the peak of $\text{Tr}\left[ \mathcal{G}\mathcal{G}^* \right]$ at $\omega_s$ almost disappears but the other peak is significantly enhanced at $\omega_h$. Consequently, due to the highly matched peak frequencies of the nanoparticle and multilayered structure, we obtain a significant thermal conductance as shown in Fig. 4(a) for $N = 21$. We thus call this effect as resonant RHT.

To interpret the behavior of $\text{Tr}\left[ \mathcal{G}\mathcal{G}^* \right]$, we show the absolute square of the $zz$ element $|G_{zz}|^2$ with respect to $k_x$ and $\omega$ in Figs. 4(c)-4(e) for the cases with the bulk and the multilayered structure with $N = 1, 21$. It is shown that, the largest value of $k_x$ for the $|G_{zz}|^2$ contour is truncated. This origins from the evanescent feature of the surface wave. The exponential $e^{i[-k_{z0}(z_2+h)+k_{z0}z_1]}$ in the GF expressed in Eq. (3) introduces a cutoff [33] at $k_\rho \approx 1/(2d)$. We see that although the largest $k_x$ of $N = 21$ is less than that of the bulk, owing to the resonance at a frequency closer to $\omega_{np}$ the conductance is much larger than that of the bulk as shown in Fig. 4(a).

We note that $f$ is assumed to be 0.5 in the above discussion. Since $f$ would change the optical properties of the multilayered structure, one can expect that it would make impact on the RHT. We find that the amplification of RHT is most prominent for $f = 0.5$ where the epsilon-near pole and epsilon-near zero frequencies coincide with each other (for more details, see Ref. [43]). In addition, we find that when $d_x > 0$, viz., a non-zero lateral distance, due to the interference between surface waves, the RHT conductance reveals an interesting oscillating and non-monotonic behavior (for more details, see Ref. [43]). This oscillation RHT is an example of an electromagnetic resonator in the context of heat radiation [29].

In conclusion, we have demonstrated that a giant resonant radiative heat transfer between nanoparticles could be achieved by inserting a multilayered structure. The physical mechanisms have been identified and elaborated to be the excitation of high-$k$ evanescent modes at the frequency close to the nanoparticle resonance. Due to this quasi-monochromaticity and resonant effect, our structure could be exploited to considerably improve the efficiency of near-field energy conversion devices.

*Acknowledgements* - This work was supported by the National Natural Science Foundation of China (Grant No. 51706053, 51776054). M. A. acknowledges support from the Institute Universitaire de France, Paris – France (UE).

———————————————

*yihongliang@hit.edu.cn
†mauro.antezza@umontpellier.fr